\documentclass[]{spie}  

\newcommand{\az}{\mathrm{az}}
\newcommand{\el}{\mathrm{el}}

\usepackage{amsmath,amsfonts,amssymb}
\usepackage{graphicx}
\usepackage[colorlinks=true, allcolors=blue]{hyperref}
\usepackage{siunitx}
\usepackage{subcaption}
\usepackage{enumitem}
\usepackage{multirow}
\usepackage{indentfirst}
\title{Data reduction pipeline for the SuMAC millimeter-wave spectrometer at the LMT}

\author[a]{A.~M.~Lapuente}
\author[b]{P.~S.~Barry}
\author[b]{M.~Becerril-Tapia}
\author[b]{C.~Benson}
\author[c,d]{C.~M.~Bradford}
\author[b]{T.~Brien}
\author[h]{S.~C.~Chapman}
\author[b]{S.~Doyle}
\author[e]{V.~G\'omez-Rivera}
\author[c]{S.~Hailey-Dunsheath}
\author[e]{J.~Hern\'andez-Aguilar}
\author[e]{J.~L.~H\'ernandez-Rebollar}
\author[e]{D.~H.~Hughes}
\author[c]{E.~Kane}
\author[a]{K.~S.~Karkare}
\author[f]{R.~McGeehan}
\author[e]{R.~Naranjo-Romero}
\author[b]{A.~Papageorgiou}
\author[g]{J.~G.~Redford}
\author[e]{I.~Rodr\'iguez-Montoya}
\author[b]{S.~Rowe}
\author[e]{D.~O.~S\'anchez-Arg\"uelles}
\author[a]{S.~Savorgnano}
\author[a]{A.~Tan}
\author[e]{J.~Vega-M\'endez}

\affil[a]{Boston University, 590 Commonwealth Ave, Boston, MA 02215, United States}
\affil[b]{Cardiff University, 5 The Parade, Cardiff, CF24 3AA, United Kingdom}
\affil[c]{California Institute of Technology, 1200 E California Blvd, Pasadena, CA 91125, United States}
\affil[d]{Jet Propulsion Laboratory, 4800 Oak Grove Drive, La Cañada Flintridge, CA 91011, United States}
\affil[e]{Instituto Nacional de Astrofísica, Óptica y Electrónica, Luis Enrique Erro 1, Santa María Tonantzintla, Puebla, CP 72480, Mexico}
\affil[f]{The University of Chicago, 5801 S Ellis Ave, Chicago, IL 60637, United States}
\affil[g]{University of California Santa Barbara, 552 University Road, Santa Barbara, CA 93106, United States}
\affil[h]{Dalhousie University, 1453 Lord Dalhousie Drive, Halifax, NS B3H 4R2, Canada}

\authorinfo{Corresponding author: Alex Lapuente (lapuente@bu.edu)}
\begin{document} 
\maketitle

\begin{abstract}
We present the data reduction and analysis pipeline for the SuperSpec-MUSCAT Collaboration (SuMAC) along with several key science data products.  In Summer 2025, SuperSpec on-chip spectrometers operating from \qtyrange[range-phrase=--,range-units=single]{190}{300}{\giga\hertz} at $R\sim200$ were deployed in the MUSCAT cryostat at the Large Millimeter Telescope, obtaining several days of on-sky data. Analysis of these data has been used to characterize noise performance and study spectra of various astrophysical sources, establishing the viability of deploying SuperSpec detectors in future instruments. These proceedings discuss the modular data reduction pipeline which uses timestreams of the kinetic inductance detectors (KIDs) and auxiliary instrument data to generate a variety of data products. We present an on-sky calibration method for converting the KID fractional frequency shift to on-sky brightness temperature. The mapmaking module generates 3D data cubes of extended sources, while the spectrum module produces a millimeter-wave spectrum of an object. The observations all incorporate measurements of the atmosphere to model and correct for absorption. We also discuss data processing choices including detector weighting, despiking, and notch filtering. Finally, we present preliminary data products including detection of carbon monoxide in NGC 253 and spectral maps of the central region of the Orion KL nebula.
\end{abstract}

\keywords{on-chip spectroscopy, kinetic inductance detectors, millimeter-wave, data reduction}

\section{Introduction}
Millimeter-wave spectroscopy provides a powerful probe of star formation and the interstellar medium across cosmic history. Within the \qtyrange[range-phrase=--,range-units = single]{190}{300}{\giga\hertz} atmospheric observing window, several atomic and molecular emission lines from dusty, star-forming galaxies can be observed. In this window, multiple rungs of the CO$(J\to J-1)$ rotational ladder can be observed across $0<z<2$. Observations at $z\sim 2$ can directly probe the cold molecular content of galaxies during the ``cosmic noon'' era of peak star formation. Line intensity mapping (LIM) experiments propose to survey emission lines such as CO$(J\to J-1)$ from many unresolved galaxies. This will allow for surveys of the universe's large-scale structure at redshifts inaccessible to traditional galaxy surveys \cite{Visbal2010,KarkareSnowmass}.

SuperSpec is an on-chip spectrometer designed for high-sensitivity measurements in the \qty{1}{\milli\meter} radio observing band \cite{Shirokoff2014}. The device uses lithographically patterned filters coupled to lumped-element titanium nitride (TiN) kinetic inductance detectors (KIDs) with typical spectral resolution of $R\sim 200$\cite{HaileyDunsheath2015}. This design allows for much greater compactness than traditional spectrometer technologies and enables instantaneous coverage of a wide range of frequencies by individual pixels.

\begin{figure}[t]
    \centering
    \begin{subfigure}[c]{0.58\textwidth}
        \centering
        \includegraphics[width=\textwidth]{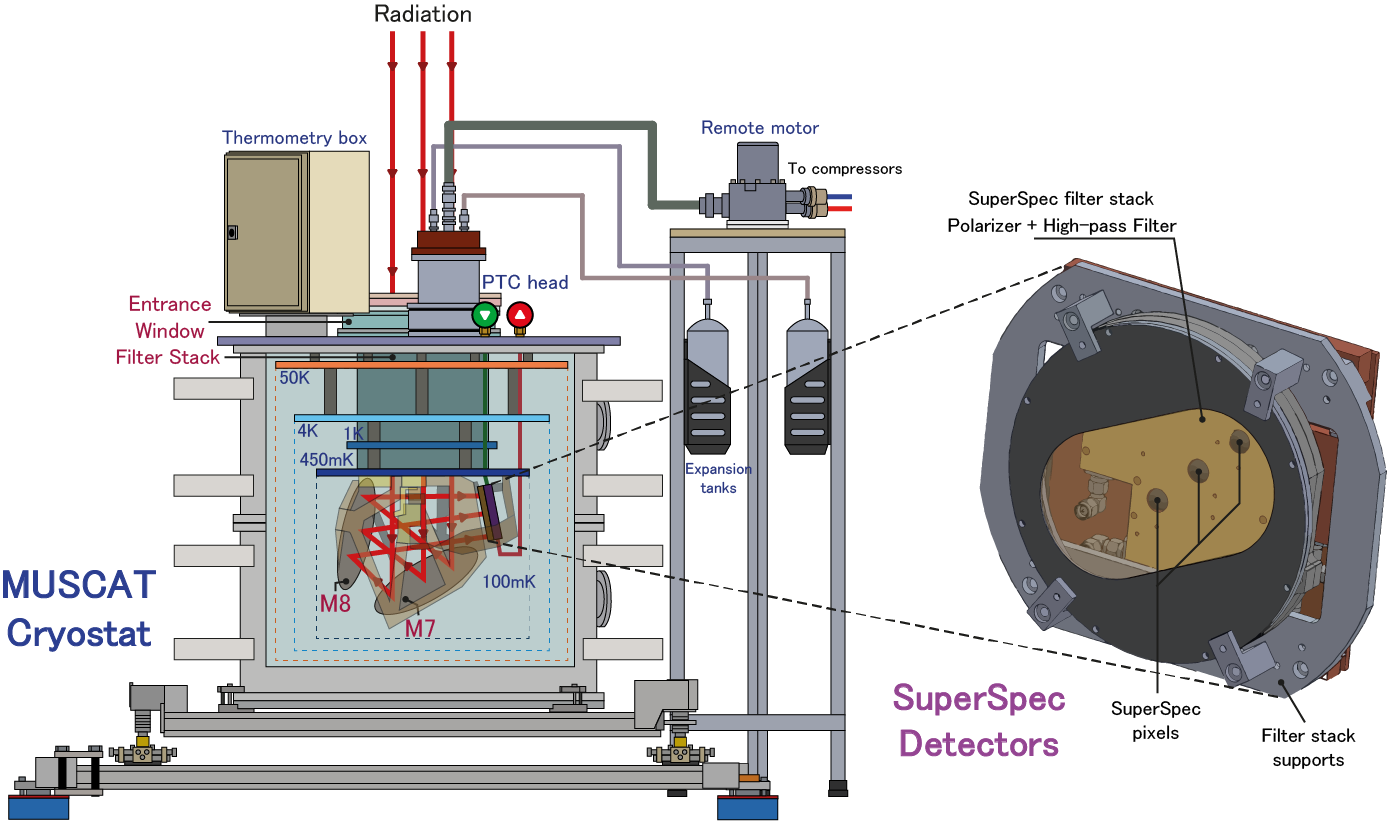}
        \caption{}
        \label{fig:sumac}
    \end{subfigure}
    \hfill
    \begin{subfigure}[c]{0.38\textwidth}
        \centering
        \includegraphics[width=\textwidth]{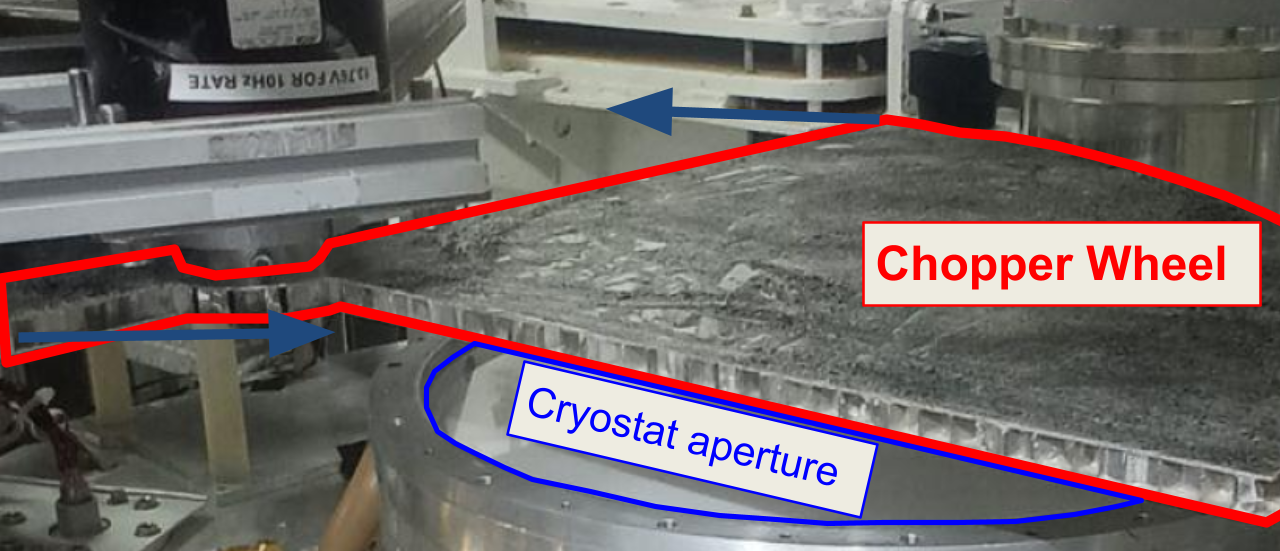}
        \caption{}
        \label{fig:chopper_wheel}
        \vspace{0.1cm}
        \includegraphics[width=\textwidth]{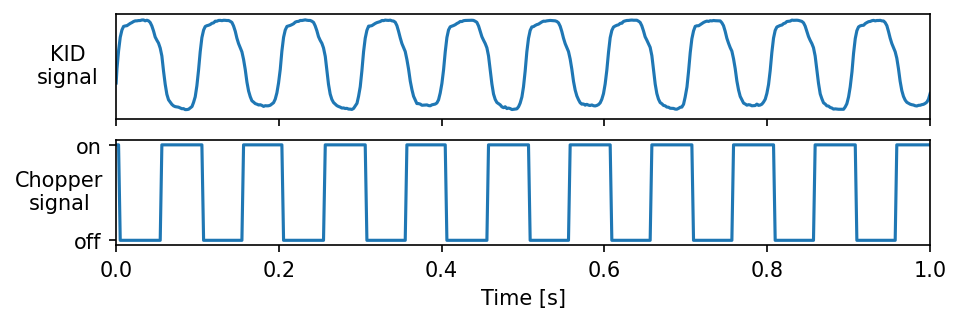}
        \caption{}
        \label{fig:chopper_sig}
    \end{subfigure}
    \caption{(a): Schematic of SuperSpec detectors within MUSCAT cryostat. Figure by M. Becerril-Tapia. (b): Chopper wheel installed over cryostat aperture. Arrows denote direction of rotation. (c): KID and chopper signals over a \qty{1}{\second} interval.}
    \label{fig:muscat_stuff}
\end{figure}

The SuperSpec-MUSCAT (SuMAC) collaboration is the joint effort to deploy and operate SuperSpec spectrometers within the MUSCAT cryostat \cite{Tapia2020,Tapia2024} (Fig.~\ref{fig:sumac}) at the Large Millimeter Telescope Alfonso Serrano (LMT). The LMT is a \qty{50}{\meter}-diameter radio telescope located at the peak of the extinct volcano Sierra Negra at an altitude of $\sim$\qty{4600}{\meter} in the Mexican state of Puebla\cite{Hughes2020}. In the Summer of 2025, the SuMAC collaboration deployed at the LMT and achieved first light. Numerous observations were conducted to characterize the instrument, calibrate the detectors, and baseline performance for future deployments. The SuMAC instrument is described in greater detail in Becerril-Tapia et al.~(2026)\cite{Tapia2026}.

\begin{figure}[ht]
    \centering
    
    \begin{subfigure}[c]{0.24\textwidth}
        \centering
        \includegraphics[width=\textwidth]{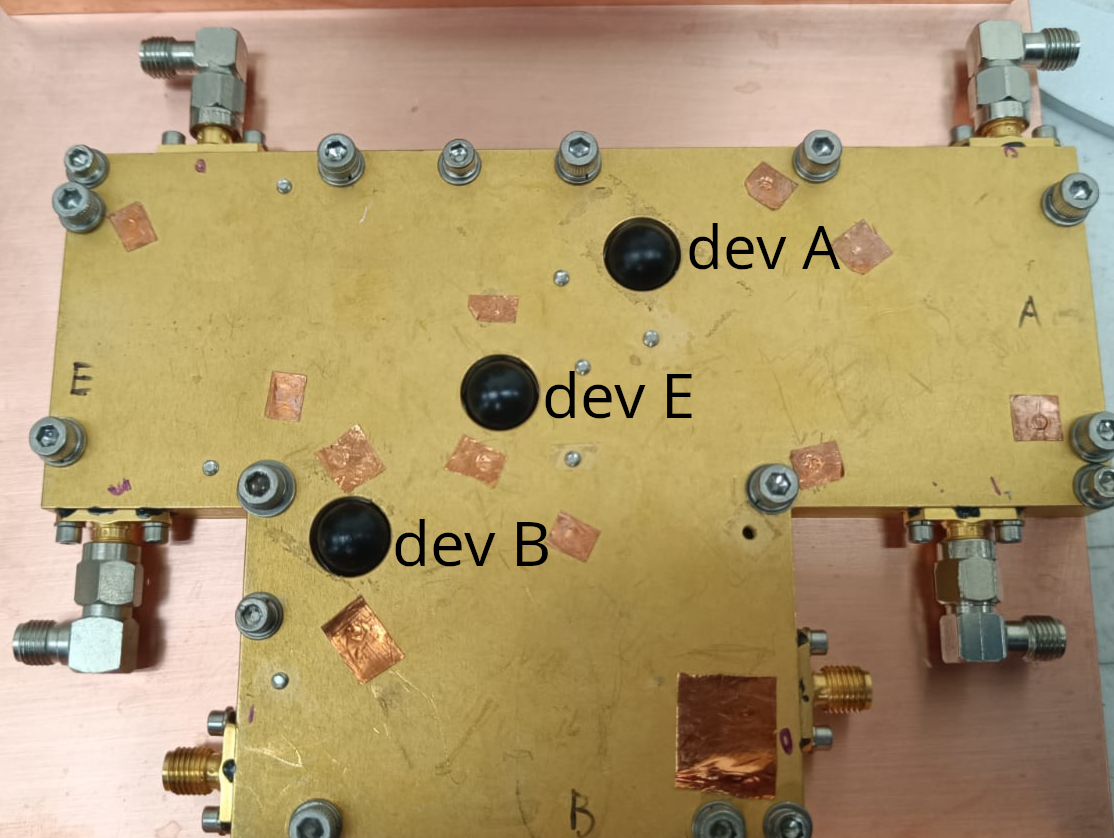}
        \caption{}
        \label{fig:device}
    \end{subfigure}
    \hfill
    \begin{subfigure}[c]{0.68\textwidth}
        \centering
        \includegraphics[width=\textwidth]{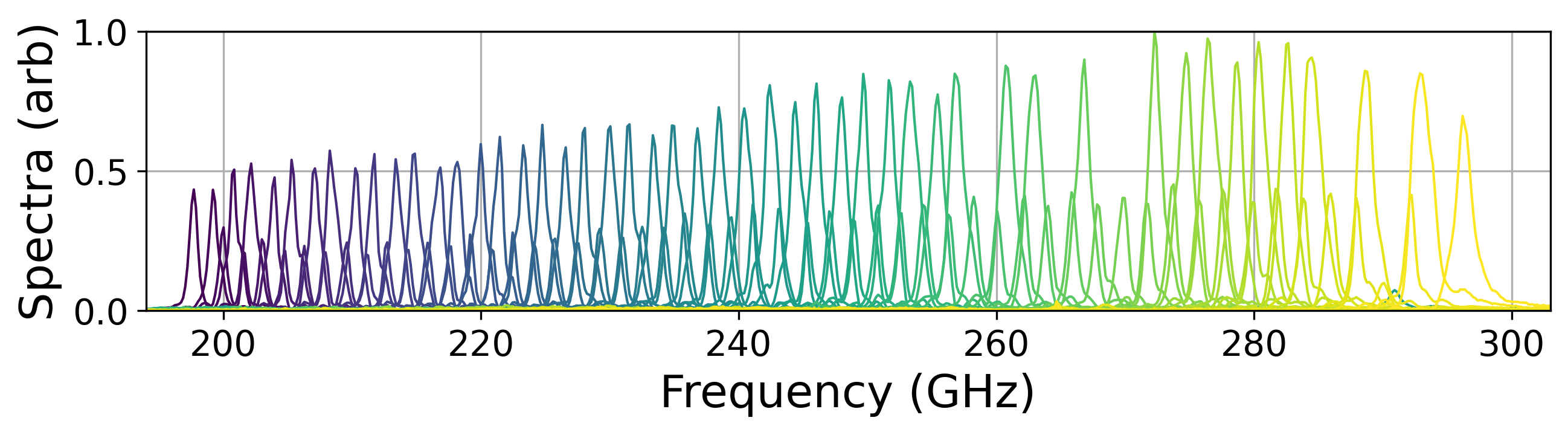}
        \caption{}
        \label{fig:filterbank}
    \end{subfigure}
    \caption{(a): The SuperSpec device. The three spectrometer chips are here labeled dev B, E, and A. (b): Filterbank of SuperSpec's central pixel. Figure by E. Kane.\cite{Kane2026}}

\end{figure}

The deployed focal plane consisted of three individual SuperSpec chips (Fig.~\ref{fig:device}). Each of these pixels is read out by a separate ROACH2 FPGA board. These proceedings will concern data obtained by the center pixel (`dev E' in Fig.~\ref{fig:device}), as spectral characterization of the side pixels is still underway. The center pixel's filterbank spans the \qtyrange[range-phrase=--,range-units = single]{190}{300}{\giga\hertz} atmospheric observing window, shown in Fig.~\ref{fig:filterbank}. Filterbank characterization is covered extensively in Kane et al.~(2026)\cite{Kane2026}.

In these proceedings, we summarize the data products acquired during our initial deployment. We describe the steps of our modularized data reduction and analysis pipeline, including how different data types are treated. We explain the atmospheric modeling routine used to estimate atmospheric opacity as well as the on-sky flux calibration routine used to calibrate our detectors. We present an initial spectrum of starburst galaxy NGC 253 showing detection of the CO$(2\to 1)$ transition line as well as raster maps of the Orion KL nebula. Finally, we summarize ongoing analysis efforts and plans for future work.

\section{Observation summary}
Between the dates of 2025-07-13 and 2025-08-29, observations of twenty-four galactic and extragalactic sources were conducted. Five solar system objects, four galactic objects, six galaxies and clusters, and nine compact AGN/quasar calibrators were observed. Data quality varied strongly from observation to observation, due to the highly variable atmospheric conditions during deployment. A small subset of these sources will be discussed in these proceedings, while other sources will be covered in future analyses. Each observation followed one of several tracking protocols to serve the broader goal of either creating maps or spectra of sources.

\begin{enumerate}
    \item Maps
    \begin{enumerate}[label=\alph*.]
        \item \textit{Standard maps}. Standard maps (SMs) are raster scans subtending $\sim\qty{100}{\arcsecond}\times\qty{100}{\arcsecond}$ about the source. These scans were conducted to characterize and calibrate the instrument (see Section \ref{sec:calibration}) and to study flux and morphological characteristics of the sources themselves.
        \item \textit{Track maps}. Track maps (TMs) are constant-elevation scans subtending $\sim\qty{100}{\arcsecond}\times \qty{1}{\arcsecond}$ about the source.
        \item \textit{Lissajous maps}. Lissajous maps (LMs) are scans of compact sources intended for beam characterization.
    \end{enumerate}
    \item Spectra
    \begin{enumerate}[label=\alph*.]
        \item \textit{Position switches}. Position switches (PS) consist of data acquired on and off a source. The telescope is first positioned such that the center pixel is centered on the source, then such that the center pixel is azimuthally offset; this alternating pattern is repeated throughout the observation. These `on-off' measurements are used to create spectra by subtracting the background from on-source values. (see Section \ref{sec:spectra}).
        \item \textit{Double position switches}. Double position switches (DPS) are similar to position switches, except that their off-source data is taken at both positive and negative azimuthal offset from the source.
    \end{enumerate}
\end{enumerate}

\section{Reduction Pipeline}

\subsection{$IQ$-$\delta f$ Conversion}
\begin{figure}[htbp]
    \centering

    \begin{minipage}{0.8\textwidth}
        \centering

        \begin{subfigure}[t]{0.48\linewidth}
            \centering
            \includegraphics[width=\linewidth]{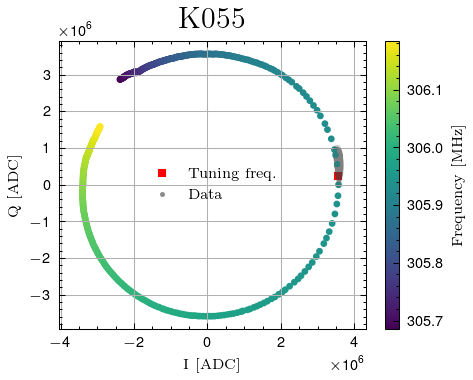}
            \caption{}
            \label{fig:good-IQ}
        \end{subfigure}
        \hfill
        \begin{subfigure}[t]{0.48\linewidth}
            \centering
            \includegraphics[width=\linewidth]{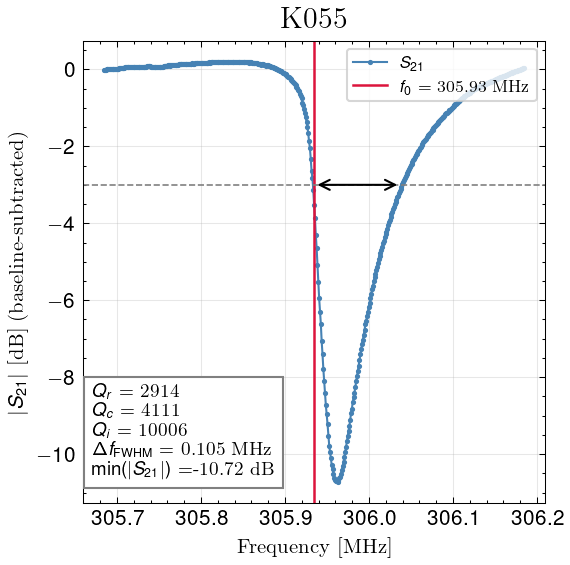}
            \caption{}
            \label{fig:S21_fitting}
        \end{subfigure}
    \end{minipage}

    \caption{(a): An IQ loop. The tuning frequency is shown by the red square, while observation data are plotted in gray. (b): The IQ loop's corresponding transmission magnitude $|S_{21}(f)|$ with derived quality factors. The resonator quality factor is first estimated as $f_{0}/\Delta f_{\rm FWHM}$. Then, the coupling quality factor $Q_{c} = \left|\frac{Q_{r}}{\min(|S_{21}|)-1}\right|$ and the internal quality factor $Q_{i} = \left(Q_{r}^{-1}-Q_{c}^{-1}\right)^{-1}$ are computed. Application of these quality factors to filterbank characterization is discussed in Kane et al.~(2026)\cite{Kane2026}.}
    \label{fig:IQ_stuff}
\end{figure}

Prior to each observation, a tuning operation is performed. Around each expected KID resonant frequency, a sweep is conducted to obtain a complex $S_{21}(f) = I(f)+iQ(f)$. The resonant frequency $f_{0}$ is then found at the point of maximum $IQ$ velocity: $f_{0} = \arg\max_{f}\left|\frac{\mathrm{d}S_{21}} {\mathrm{d}f}\right|$. Then, the sweep is fit to a circle, $(I-I_{c})^{2}+(Q-Q_{c})^{2}=r^{2},$ to obtain the center $(I_{c},Q_{c})$ and the radius $r$. To account for cable delay, the circle is then rotated by an angle $\theta = \tan^{-1}\left(\frac{Q_{0}-Q_{c}}{I_{0}-I_{c}}\right)$, where $(I_{0},Q_{0}) = S_{21}(f_{0})$. We thus obtain a centered and derotated $S_{21}$, $S_{21}' = (S_{21}-S_{c})e^{-i\theta}$ (Fig.~\ref{fig:IQ_stuff}). From $S_{21}'$, we obtain a phase,
\begin{equation}
    \phi(f) = \tan^{-1}\left[\frac{\mathfrak{Im}\left((S_{21}(f)-S_{c})e^{-i\theta}\right)}{\mathfrak{Re}\left( (S_{21}(f)-S_{c})e^{-i\theta}\right)}\right]= \tan^{-1}\left(\frac{Q'(f)}{I'(f)}\right)\,.
\end{equation}
This empirical mapping is inverted and interpolated to obtain $f(\phi)$. When data are later taken, this is used to convert each time-ordered data point $S_{21}(t)$ to a frequency $f(t)$. In turn, this is used to calculate a frequency shift $\delta f=f-f_{0}$ and a fractional frequency shift, $\delta f/f_{0}$.

If a resonance is driven with too much readout power, the detector may become overdriven\cite{Swenson2013}. This results in a discontinuity in the $IQ$ loop, meaning the phase cannot be accurately recovered. For each observation, overdriven detectors are ruled out by inspection and excluded from further analysis. In some cases, especially when the chopper wheel is used (see Section \ref{sec:chopper}), the time-ordered data may depart from the region swept out by the $IQ$ loop. Likewise in this case does phase calculation fail, and data from these detectors too must be discarded.

\subsection{Chopper demodulation}\label{sec:chopper}
For a subset of observations, a rotating chopper wheel was placed over the cryostat aperture, alternating between viewing the sky and reflecting back into the cryostat (Fig.~\ref{fig:chopper_wheel}). It rotated with a chop frequency of $\sim\qty{10}{\hertz}$ to modulate timestreams out of the region dominated by $1/f$ detector noise. An optical sensor was installed to record the chopper wheel's position (Fig.~\ref{fig:chopper_sig}). Modeling the chopper signal as a square wave, we use an IQ demodulation routine to separate the observed KID signal into two components phase-separated by $\qty{90}{\degree}$. To recover the on-sky signal, we replace the chopped data with the timestream of the on-sky phase component. 

\subsection{Timestream filtering}\label{sec:filtering}
\begin{figure}
    \centering
    \includegraphics[width=1.0\linewidth]{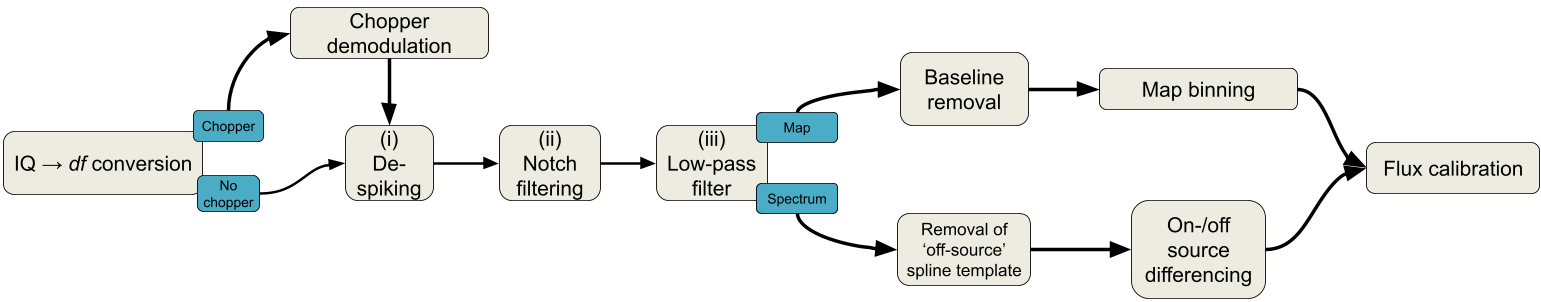}
    \vspace{5pt}
    \caption{Flowchart representing modularized data reduction from raw data to final products.}
    \label{fig:reduction_flowchart}
\end{figure}

At this stage, we employ a series of data cleaning steps. All data products are subject to the first three steps, while later steps depend on the type of observation being filtered. This process is illustrated in Fig.~\ref{fig:reduction_flowchart}.

\begin{enumerate}[label=\roman*.]
    \item \textit{Despiking}. This step removes spike-like abnormalities caused by cosmic rays or readout glitches. For the entire timestream data vector $d$, the gradient $\nabla d$ is computed, along with $\mu = \mathrm{median}(\nabla d)$ and $\sigma = \mathrm{std}(\nabla d)$. A spike is detected at the $i^{\rm th}$ index if $|\nabla d_{i} - \mu|> 4\sigma$. Local statistics $\mu_{W}$ and $\sigma_{W}$ are then computed on a masked interval $W = [i-50:i-25]\cup[i+25:i+50]$. If $|\nabla d_{i}-\mu_{W}|>4\sigma_{W}$, then the spike point is replaced with $d_{i}\sim \mathcal{N}\left(\mu_{W},\sigma_{W}^{2}\right)$. This algorithm continues until no spikes remain.  In both TM and (D)PS observations, a median detector has $\sim 0.02\%$ of its data vector replaced in the despiking step.
    \item \textit{Notch filtering}. This is a Fourier-domain filter which attenuates signals in narrow regions around \qtylist[list-units = single]{11.4;15}{\hertz} to mitigate vibrational noise induced by the cryogenic system. These frequencies were previously identified by MUSCAT as resulting from mechanical pumps within the cryostat\cite{TapiaThesis}.
    \item \textit{Low-pass}. This is a Fourier-domain filter which convolves the timestream with a $\mathrm{sinc}$ function to attenuate signals above $f_{s}/N$, where $f_{s}=\qty{488}{\hertz}$ is the readout sampling frequency and $N$ is an integer. $N$ can vary depending on the source's angular extent and is chosen empirically such that the source's signal-to-noise ratio is maximized. For point-like sources, $N\sim 100$, while for extended sources, $N\sim 30$. A Blackman window is applied to smooth the transition and reduce ringing. To mitigate edge effects incurred by the $\mathrm{sinc}$ filtering,  the first and last $500$ datapoints ($\sim \qty{1.02}{\second}$) are removed.
\end{enumerate}
Map-like observations are then subject to two additional steps.
\begin{enumerate}[label=\alph*.]
    \item \textit{Baseline removal}. This step uses a moving Savitzky-Golay filter of order $3$ and a window size of 151 samples ($\sim \qty{0.309}{\second}$) to account for long-term baseline drift. This drift is likely due to a combination of atmospheric and instrumental fluctuations.
    \item \textit{Map binning}. We use recorded $(\az,\el)$ or computed (RA,Dec) data to bin the timestream into spatial coordinates, weighted by the number of hits per bin, $N_{i,j}$:
\begin{equation}\label{eq:binning}
    \frac{\delta f}{f_{0}}\Bigg|_{[i,j]} = \frac{\sum_{[\az,\el]\in [i,j]}d(\az,\el)}{N_{i,j}}\,.
\end{equation}
\end{enumerate}

\noindent Spectrum-like observations are also subject to two additional steps.
\begin{enumerate}[label=\alph*.]
    \item \textit{Off-source spline removal}. Using pointing data, sections of the timestream are tagged as `on' or `off' source. Each off-source segment is divided into subregions, with the median detector response of each region calculated. A smoothing spline is then fitted to these median values and evaluated over the entire observation. This baseline is then subtracted from the full timestream to account for long-term baseline drift.
    \item \textit{On-off differencing}. For each pair of `on'/`off' segments, the off-source signal is subtracted from the preceding on-source signal to remove residual sky and instrumental backgrounds. The mean and standard error of the mean are recorded, and segments are combined using inverse-variance weighting.
\end{enumerate}

\section{Atmosphere}\label{sec:atmosphere}

For much of the season, atmospheric conditions varied significantly and thus require a significant degree of care. Discussions of on-sky noise performance can be found in Redford et al.~(2026)\cite{Redford2026} and Savorgnano et al.~(2026)\cite{Savorgnano26}; the former offers a comparison to lab noise analyses, while the latter focuses on noise performance during LIM-style observations. A zenith-pointing water vapor radiometer at the LMT typically records the opacity at \qty{225}{\giga\hertz} ($\tau_{225}$). However, the radiometer ceased functioning midway through deployment; for affected observations, we must estimate opacity through other means. We employ an iterative approach integrating NASA MERRA-2 meteorological data and the \texttt{am} atmospheric modeling software\cite{am}. 

We make use of three-dimensional assimilated pressure-level data (M2I3NPASM, v5.12.4)\cite{MERRA2}, provided in three-hour intervals over a longitude$\times$latitude grid with spacing $\qty{0.625}{\degree}\times \qty{0.5}{\degree}$. However, we must account for atmospheric variation below this timescale and spatial resolution. We generate \texttt{am} configuration files from  MERRA-2\footnote{Scott Paine, personal communication}, interpolating data from the spatial grid of MERRA-2 cells to estimate atmospheric parameters at the telescope's precise coordinates. We account for the LMT's altitude, including only data from atmospheric levels above the Sierra Negra's peak. The output \texttt{am} configuration files consist of atmospheric layers from the troposphere through the mesosphere and ascribe to each layer a pressure, temperature, H$_{2}$O mixing ratio, and O$_{3}$ mixing ratio. Each file contains an \texttt{Nscale} argument, which parametrizes the atmosphere's precipitable water vapor by rescaling the H$_{2}$O column density.

\begin{figure}[h!]
    \centering
    \begin{subfigure}{0.48\textwidth}
        \centering
        \includegraphics[width=\linewidth]{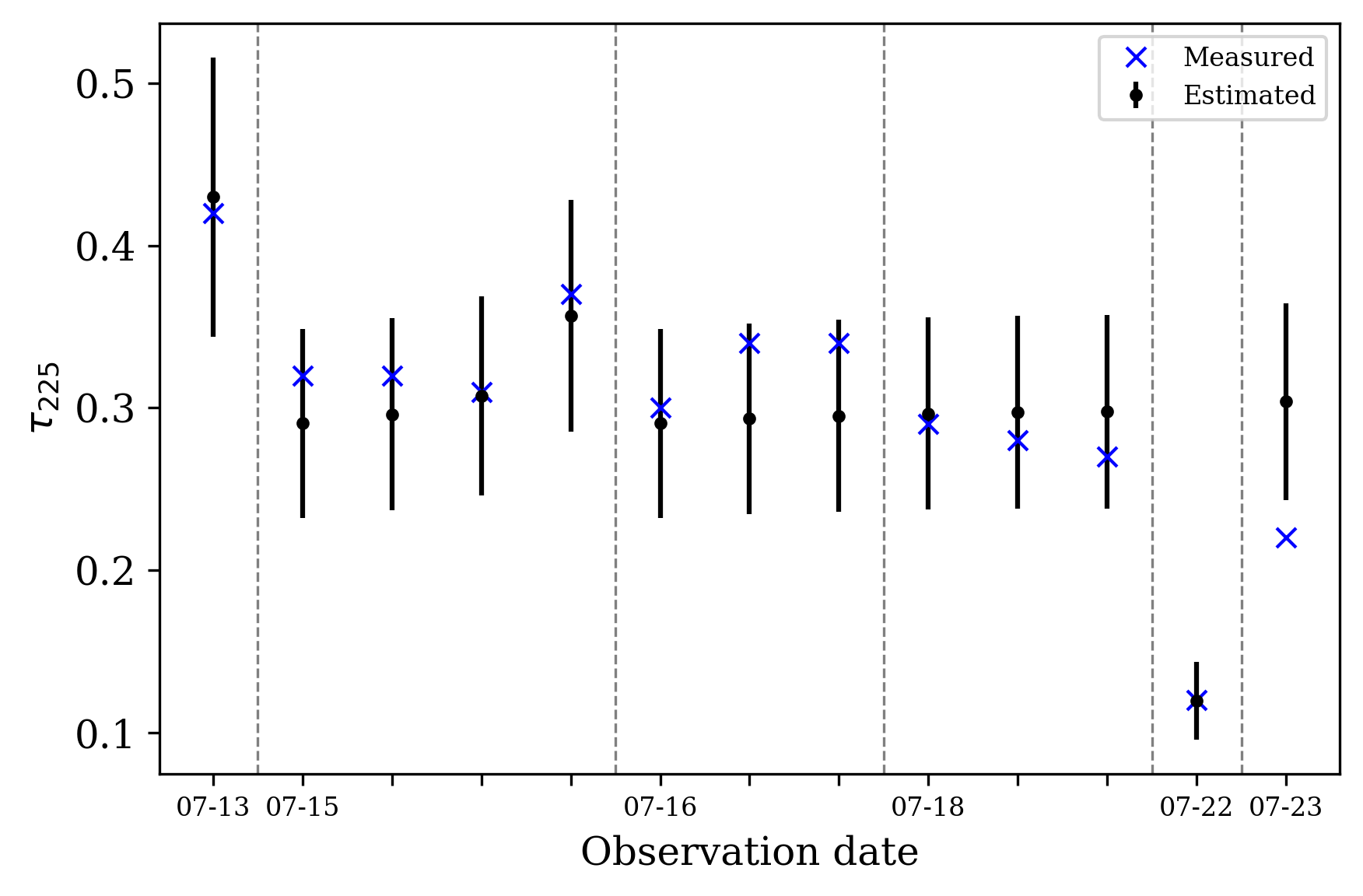}
        \caption{}
        \label{fig:tau_est}
    \end{subfigure}
    \hfill
    \begin{subfigure}{0.48\textwidth}
        \centering
        \includegraphics[width=\linewidth]{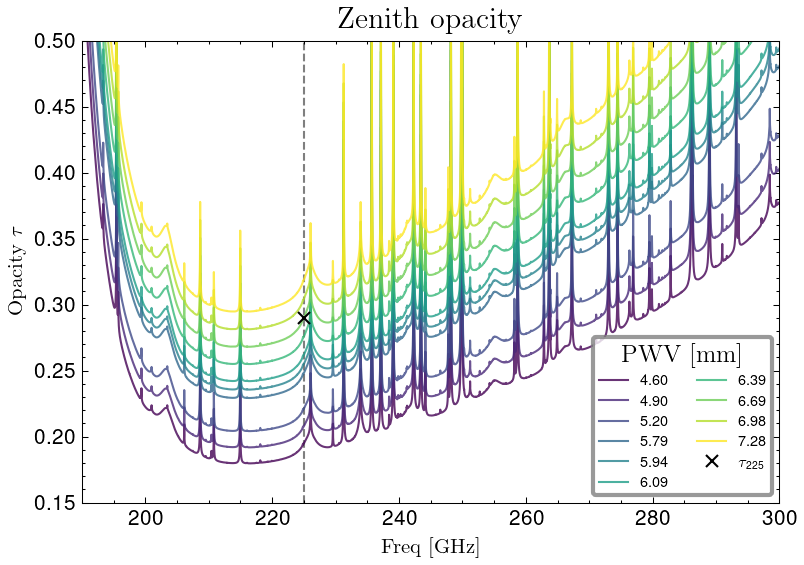}
        \caption{}
        \label{fig:tau_iter}
    \end{subfigure}

    \caption{(a): Validation of $\tau_{225}$ estimation method using extant radiometer data. An uncertainty of $20\%$ is shown. (b): Iterative method for obtaining $\tau(\nu)$. Here, the optimal spectrum corresponds to a PWV of \qty{6.69}{\milli\meter} as it closely matches the estimated $\tau_{225}$.}
    \label{fig:am_stuff}
\end{figure}

For an observation conducted at time $t_{0}$, we begin by generating a $\tau(\nu)$ spectrum at the three-hour intervals $t_{i}$ and $t_{f}$ before and after $t_{0}$ (respectively). These spectra give us $\tau_{225,i}$ and $\tau_{225,f}$. We approximate that opacity changes linearly over these timescales and thus obtain
\begin{equation}\label{eq:tau_est}
    \tau_{225,0} = \tau_{225,i}  + \frac{\tau_{225,f}-\tau_{225,i}}{t_{f}-t_{i}}(t_{0}-t_{i})\,.
\end{equation}
We validate this approximation method against extant radiometer data and find good agreement (Fig.~\ref{fig:tau_est}). For downstream error estimation, we conservatively estimate $\delta \tau(\nu_{k})/\tau(\nu_{k})=0.2$.

We now employ an iterative process to generate a $\tau(\nu)$ spectrum from this interpolated $\tau_{225,0}$ value. We begin by guessing a value \texttt{Nscale}$_{0}$ and generating a grid of \texttt{Nscale} values about this fiducial value. We then use \texttt{am} to generate opacity spectra $\tau_{\tt am}(\nu)$ over this grid. We select the value \texttt{Nscale}$_{1}$ that minimizes $|\tau_{225}-\tau_{\tt am}(\qty{225}{\giga\hertz})|$. We then repeat this process over a finer grid centered at \texttt{Nscale}$_{1}$. This procedure ends when we find a value \texttt{Nscale}$_{*}$ such that $|\tau_{225}-\tau_{\tt am}(\qty{225}{\giga\hertz})|\leq 0.01$. An example of this iterative procedure is illustrated in Fig.~\ref{fig:tau_iter}.

\section{Flux Calibration} \label{sec:calibration}
\begin{figure}[hb]
    \centering
    \begin{subfigure}{0.6\textwidth}
        \centering
        \includegraphics[width=\linewidth]{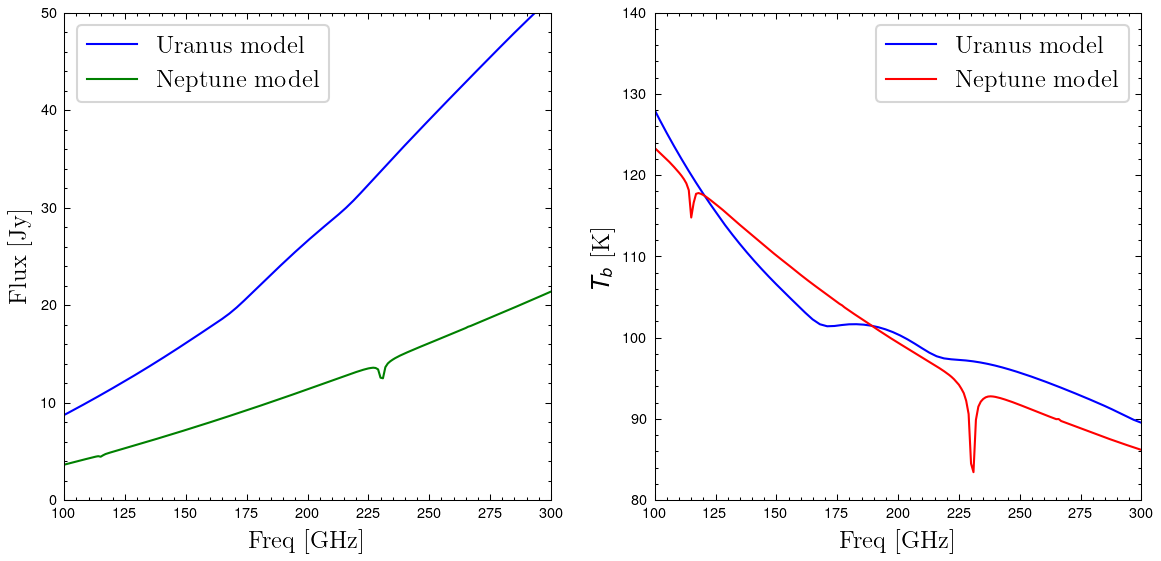}
        \caption{}
        \label{fig:esa_models}
    \end{subfigure}
    \hfill
    \begin{subfigure}{0.36\textwidth}
        \centering
        \includegraphics[width=\linewidth]{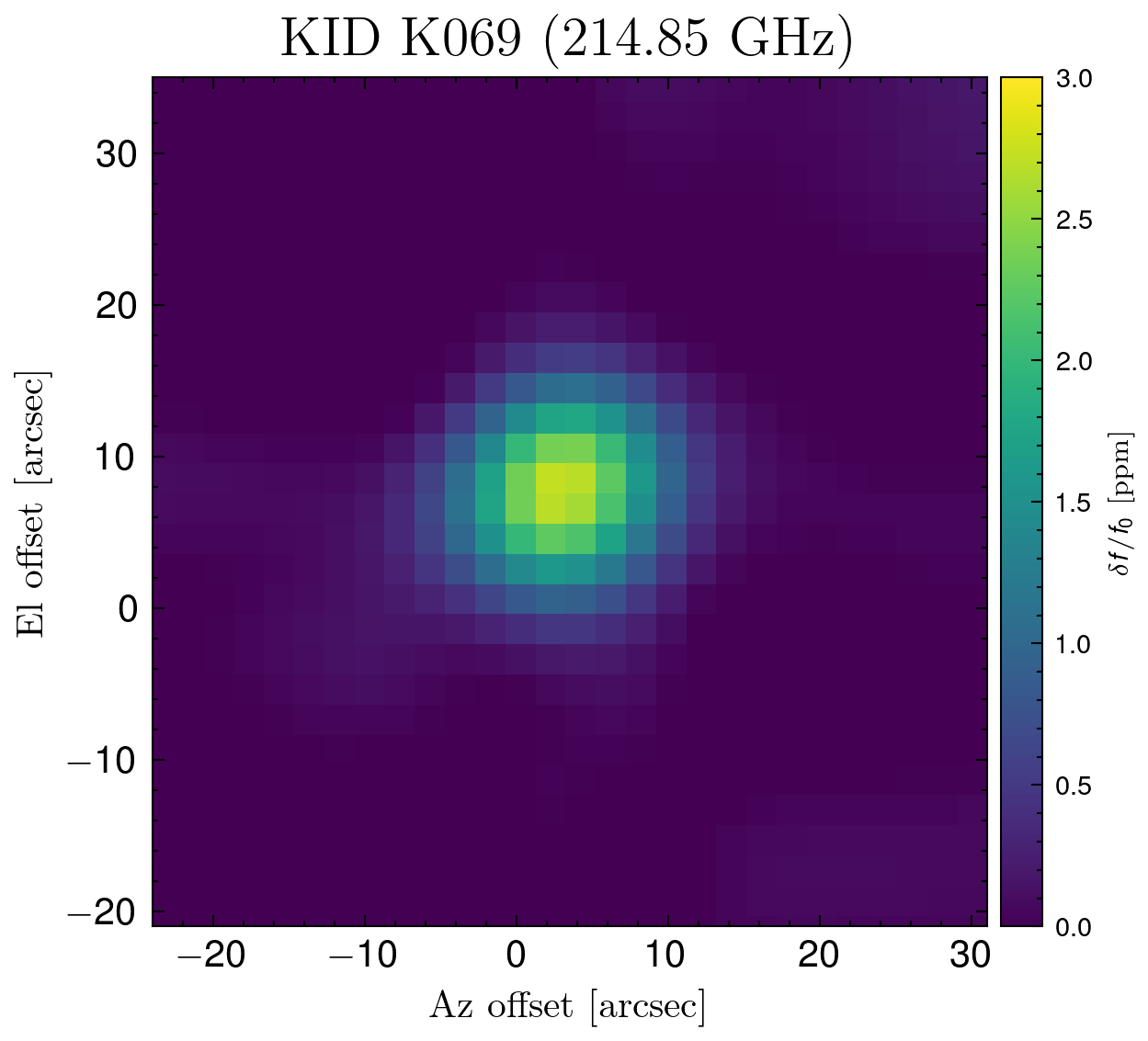}
        \caption{}
        \label{fig:neptune_beammap}
    \end{subfigure}

    \caption{(a): ESA4 and ESA5 flux and brightness temperature models for Uranus and Neptune, respectively. (b): Map of Neptune used to calibrate a detector.}
    \label{fig:flux_modeling}
\end{figure}

In order to convert each KID's fractional frequency shift into physical units, we employ flux calibration using maps of Uranus and Neptune (Fig.~\ref{fig:neptune_beammap}).

\subsection{Determining an offset}
For each map used in flux calibration, we first determine the source's location relative to telescope pointing. We first perform a preliminary unweighted co-add across optical KIDs to obtain $\bar{d}(\mathrm{az},\mathrm{el}) = \frac{1}{N}\sum_{k}^{N}d_{k}(\mathrm{az},\mathrm{el})$, where $k$ indexes optical KIDs and $d(\mathrm{az},\mathrm{el})$ is the fractional frequency shift at a given point. We then estimate a source location by finding $(\mathrm{az}_{*},\mathrm{el}_{*}) = \arg\max_{\mathrm{az},\mathrm{el}}\bar{d}(\mathrm{az},\mathrm{el})$. Since the angular radius of our calibrators is $\sim \qtyrange[range-phrase = -]{1}{3}{\arcsecond}$, we define a circular fitting region $R$ given by $R =\left \{(\az,\el) | (\az-\az_{*})^{2} + (\el-\el_{*})^{2} \leq (\qty{5}{\arcsecond})^{2}\right\}$ to encompass the source along with the background. We obtain the root-mean-square (RMS) noise $N_{k}$ outside this region to obtain normalized weights, $w_{k} = \frac{N_{k}^{-2}}{\sum_{j}N_{j}^{-2}}$. Using these, we perform a weighted coadd to obtain $\tilde{d}(\az,\el) = \sum_{k}^{N}w_{k}d_{k}(\az,\el)$. We model the beam as a Gaussian and the planet as a circle; we use a least-squares algorithm to obtain best fit parameters $\{\az_{0},\el_{0},\sigma,A,C\}$. Since we have obtained these parameters by combining data from uncalibrated detectors, the only parameters of interest so far are the source coordinates, $\{\az_{0},\el_{0}\}$.

\subsection{Individual KID Fitting}
Having now obtained a reliable location for the source, we apply the same fitting procedure to each KID's timestream, except with $(\az_{0},\el_{0})$ fixed. We thus obtain for each KID $k$ a set of best-fit parameters $\{A_{k},\sigma_{k},C_{k}\}$. $A_{k}$ characterizes the amplitude, $\sigma_{k}$ characterizes the beam's intrinsic width, and $C_{k}$ characterizes a constant offset from zero. For all but the least responsive KIDs, $C_{k}\ll A_{k}$.

\subsection{Source modeling}
For both planetary calibrators, we use models developed for the calibration of the SPIRE instrument. For Uranus, we use the `ESA4' dataset\cite{Orton2014,OrtonESA4} and for Neptune, we use the `ESA5' dataset\cite{ESA5}. These datasets provide flux $F_{\nu}$ and brightness temperature $T_{\rm b}$ over our frequency range (Fig.~\ref{fig:esa_models}).

We account for atmospheric extinction via the term $e^{-\tau_{\rm cal}(\nu_{k})X_{\rm cal}}$, where $\tau(\nu_{k})$ is the opacity at detector $k$'s center frequency (calculated as in Section \ref{sec:atmosphere}) and $X_{\rm cal}$ is the airmass, given by the cosecant of the source's elevation. For each calibration measurement, we thus find for each detector a gain factor $G_{k} = \frac{T_{b}(\nu_{k})}{A_{k}}e^{-\tau_{\rm cal}(\nu_{k})X_{\rm cal}}$. The relative uncertainty for each gain factor is
\begin{equation}\label{eq:cal_unc}
    \frac{\delta G_{k}}{G_{k}} = \sqrt{ \left[\frac{\delta T_{b}(\nu_{k})}{T_{b}(\nu_{k})}\right]^{2} + \left[\frac{\delta A_{k}}{A_{k}}\right]^{2} + \left[\delta \tau_{\rm cal}(\nu_{k})X_{\rm cal}\right]^{2}}.
\end{equation}

Both SPIRE calibrator models report a peak-to-peak flux uncertainty of $5\%$, which we treat as a symmetric uncertainty of $\pm 2.5\%$. In the Rayleigh-Jeans regime ($h\nu \ll k_{\rm B} T_{b}$), flux density is proportional to brightness temperature, so we can set $\delta T_{b}(\nu_{k})/T_{b}(\nu_{k}) = 0.025$. The least-squares fitting algorithm yields a typical relative uncertainty on the amplitude $A_{k}$ of $\delta A_{k}/A_{k}\sim 0.1$. To confirm this was a reasonable estimate of uncertainty, a more robust Markov Chain Monte Carlo (MCMC) fitting algorithm was performed using \texttt{emcee}\cite{Foreman-Mackey2013} to find fit parameters for a subset of calibration maps; good agreement was found between the two methods. For physical atmosphere values (i.e., $X\geq 1$), calibration uncertainty is dominated by the $\delta\tau_{\rm cal}(\nu_{k})X_{\rm cal}$ term.

We then average over valid gain factors to obtain $\overline{G}_{k}$ for each detector, propagating the uncertainty from each individual gain factor. When applying these gain factors to an observation, we again account for atmospheric extinction by applying $e^{+\tau_{\rm obs}(\nu_{k})X_{\rm obs}}$. This accounts for differences in opacity and telescope elevation between calibration and `science' measurements.

\section{Example spectrum -- NGC 253} \label{sec:spectra}
\begin{figure}[ht]
    \centering
    \includegraphics[width=1\linewidth]{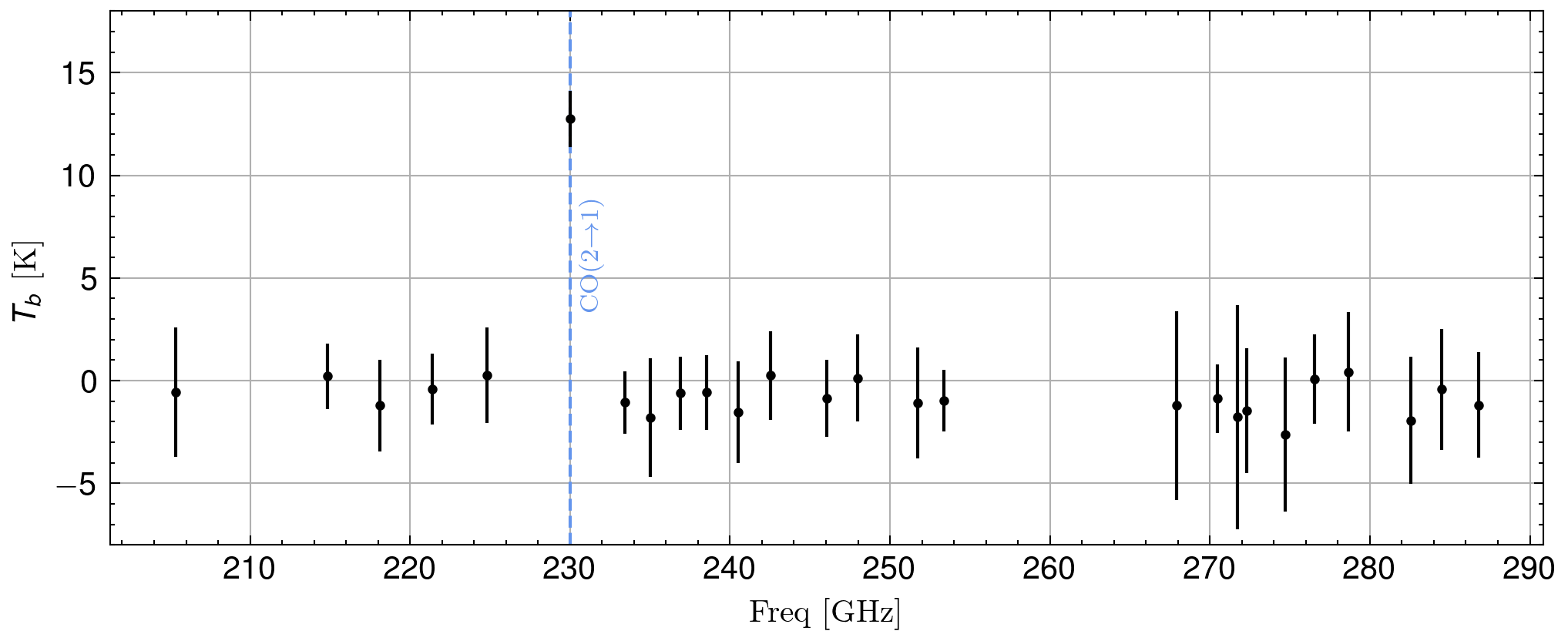}
    \caption{Example preliminary spectrum of starburst galaxy NGC 253.}
    \label{fig:spectrum}
\end{figure}

NGC 253 (or the Sculptor Galaxy) is a starburst galaxy at a distance of 3.5 Mpc. As its rate of star formation is high, it is rich in molecular emission lines including the CO$(2\to 1)$ rotational transition line at $\qty{230.538}{\giga\hertz}$. We present a preliminary spectrum (Fig.~\ref{fig:spectrum}) acquired from a single DPS observation with a 50\% chopper duty cycle. The rotational transition line can clearly be seen and is recorded at \qty{12.8+-1.3}{\kelvin}. This spectrum consists only of data from detectors which passed a series of stringent, conservative quality cuts intended to remove overdriven, unresponsive, or severely noise-dominated detectors. Uncertainties here are derived from the scatter among individual `on-off' segments and do not reflect photon-limited noise performance; our best estimate of instrument sensitivity is discussed at length in Redford et al.~(2026)\cite{Redford2026}. Instead, this scatter results from residual drift that the current spline-subtraction method (described in Section \ref{sec:filtering}) fails to fully capture. Planned improvements to this background subtraction method are detailed in Section \ref{sec:outlook}.

\section{Example map -- Orion KL Nebula}
\begin{figure}[ht]
    \centering

    \includegraphics[width=1\linewidth]{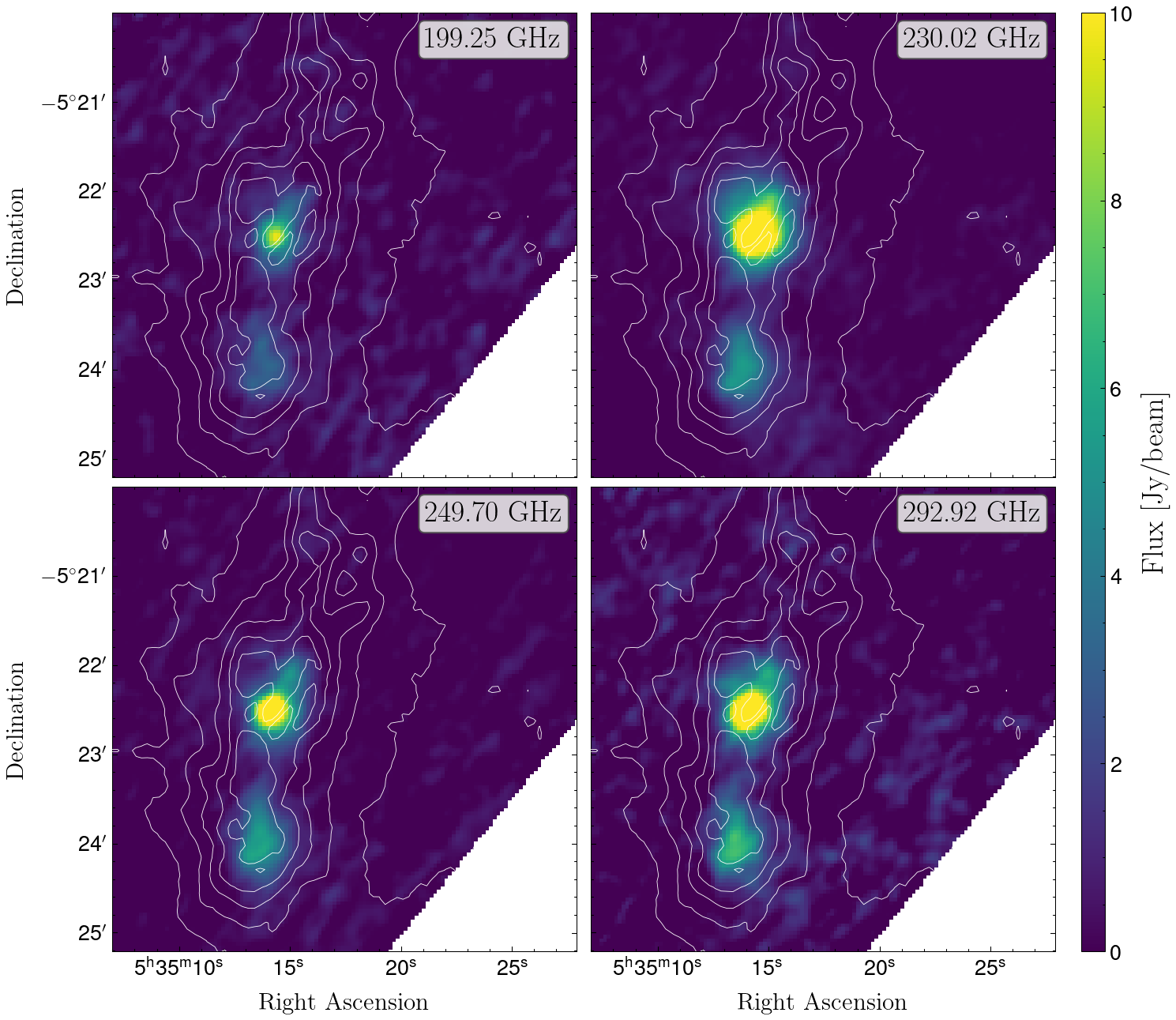}    \

    \caption{The Orion KL nebula at four frequencies across our observing bandwidth. The nebula's hot core can be seen, as well as its fainter southern lobe. White contours show the JCMT SCUBA-2 \qty{850}{\micro\meter} continuum emission at significance levels of $50$, $100$, $150$, $200$, and $250\sigma$, where $\sigma$ is the local RMS noise derived from the map's per-pixel variance\cite{WardThompson07,Kirk18}.}
    \label{fig:okl_maps}
\end{figure}

The Orion KL (or Kleinmann-Low) nebula is a star-forming region located within the Milky Way galaxy. It exhibits a warm dust continuum, making it bright in the millimeter-wave regime\cite{Wright85}. Here, we present example maps (Fig.~\ref{fig:okl_maps}) of the nebula obtained by coadding twelve observations acquired over a four-day period during August of 2025. The source's extended morphology can be seen, and its flux density is shown to vary with frequency.

\newpage
\section{Outlook and Future Plans}\label{sec:outlook}
Due to the quantity and quality of data acquired during our initial observing run, analysis of data products is ongoing. We are currently implementing improvements to the analysis routine which a future publication will describe in further detail. Among these improvements are:
\begin{enumerate}
    \item \textit{Spectral characterization of additional pixel}. Only the center pixel has been spectrally characterized with a Fourier transform spectrometer (FTS) in its deployment configuration. We are currently performing FTS measurements on one of the side pixels (`dev B' in Fig.~\ref{fig:device}), doubling the number of spectral channels accessible in our data.
    \item \textit{Improved modeling of chopper signal}. As explained in Section \ref{sec:chopper}, we currently model the chopper signal as a square wave. However, the chopper takes a nonzero amount of time to fully (un)cover the aperture (as evinced by the KID signal in Fig.~\ref{fig:chopper_sig}). We plan to use a more accurate, physical model of the chopper blade geometry to update our demodulation algorithm. This will improve our ability to fully recover the on-sky signal component.
    \item \textit{Improved characterization and removal of long-term drift}. The current methods of removing long-term background drift (Savitzky-Golay filtering and spline fitting; see Section \ref{sec:filtering}) are generally effective but imperfect. In mapping extended sources with diffuse regions (such as the Orion KL nebula), na\"ive baseline subtraction runs the risk of washing out faint, large-scale structure. In the case of position-switching observations, there are significant fluctuations on short timescales not fully captured by the spline model.

    The long-term baseline drifts we observe could be due to multiple sources, including atmospheric fluctuations, focal plane temperature, or readout electronics. Each of these effects has a different signature in the correlation structure across detectors. To study and more thoroughly account for them, we are investigating a common mode removal routine to isolate and remove common noise sources.

    \item \textit{Analysis of additional sources}. Beyond the example of the Orion KL nebula included in these proceedings, several other extended sources were observed. Future analyses will include high-resolution images and morphological characterization of these sources.
\end{enumerate}

\acknowledgments 
Alex Lapuente's attendance of the 2026 SPIE Astronomical Telescopes + Instrumentation conference was supported in part by a Boston University Photonics Center Student Travel Support Grant.

\bibliography{report} 
\bibliographystyle{spiebib} 

\end{document}